\documentclass[english]{article}
\usepackage[T1]{fontenc}
\usepackage[latin1]{inputenc}
\usepackage{geometry}
\usepackage{verbatim}
\usepackage{float}
\usepackage{graphicx}

\makeatletter

\newcommand{\lyxaddress}[1]{
\par {\raggedright #1
\vspace{1.4em}
\noindent\par}
}

\usepackage{babel}
\makeatother
\begin{document}

\title{Design of quasi-symplectic propagators for Langevin dynamics}

\author{Simone Melchionna}

\maketitle

\lyxaddress{\begin{center}
SOFT-INFM-CNR, Department of Physics, \\University of Rome La Sapienza,
\\P.le A.Moro 2, 00185, Rome, Italy
\par\end{center}}

Correspondence: simone.melchionna@roma1.infn.it

\begin{center}
\today
\par\end{center}

\begin{abstract}
A vector field splitting approach is discussed for the systematic
derivation of numerical propagators for deterministic dynamics. Based
on the formalism, a class of numerical integrators for Langevin dynamics
are presented for single and multiple timestep algorithms. 
\end{abstract}
\newcommand{\fld}{{V}}

\newcommand{\LL}{{\cal L}}

\newcommand{\UU}{{\cal U}}

\renewcommand{\SS}{{\cal S}}

\newcommand{\NN}{{\cal N}}

\newpage

\section{Introduction}

The design of efficient and stable propagators is a central issue
in the numerical study of classical and quantum systems. The way the
state is propagated in time affects the quality of the simulated trajectory
and static and dynamical ensemble averages, as much as the capacity
to efficiently sample the available phase space.

Molecular Dynamics (MD) is the reference technique for the study of
condensed matter systems \cite{frenkel-smit}. Although, in principle,
time discretization introduces an error which affects the statistics
of the simulated systems, from the point of view of numerical control,
MD offers a number of advantages. The underlying conservative and
time-reversible nature of the Hamiltonian equations allows to monitor
the quality of the dynamics as a function of the time step and the
implemented interaction potential. Despite energy conservation is
never attained in practice, keeping the energy fluctuations small
is a necessary, although not sufficient, condition to avoid statistical
bias in the computed averages. 

The lack of drifts in the energy has long been recognized as a key
requirement to control long-time stability. In mechanical terms, long-time
stability follows from the symplectic nature of the propagator, a
distinguishing feature of Hamiltonian dynamics \cite{leimkuhler}.
In statistical terms, symplecticity implies exact conservation of
measure in phase space, i.e. the possibility of applying Gibbs statistical
mechanics to a well-defined and conserved ensemble of physical states.
Moreover, symplecticity and time-reversibility are the basis to combine
MD with Monte Carlo, in the so-called Hybrid Monte Carlo method \cite{duane},
where the acceptance test removes any systematic bias due to the usage
of a large timestep. Another important benefit of conservative dynamics
is the possibility to employ ad-hoc quasi-Hamiltonian dynamics to
sample ensembles different from the microcanonical, a popular choice
being the Nosè-Hoover dynamics \cite{frenkel-smit} by retaining the
same pleasant features of the energy conserving dynamics. 

One of the most popular numerical propagators originates from the
pioneering work of Verlet \cite{verlet} who employed an algorithm
first introduced by St\"ormer \cite{hairer}. The distinguishing
features of the Verlet (also known as St\"ormer-Verlet) algorithm
are its simplicity, as compared to predictor-corrector or Runge-Kutta
methods \cite{leimkuhler}, time-reversibility and the lack of numerical
drifts. Therefore, in spite of its limited accuracy, being quadratic
in the timestep, the Verlet propagator stands as the reference algorithm
for MD. Some years after its discovery, a systematic derivation based
on an operatorial splitting approach was proposed \cite{suzuki}.
By considering the case of a separable Hamiltonian, the so-called
Trotter factorization \cite{trotter} was applied to derive the Verlet
algorithm and its equivalents, in particular the Velocity Verlet (VV)
version. A crucial benefit of such operatorial approach is its generalization
to a multiple time step (MTS) scheme \cite{respa}, which extends
the single time step (STS) one. In the MTS, the advance in time of
the phase space state follows the temporal evolution of forces in
an asyncronous way, i.e. by treating fast and slow interactions at
different levels. The efficiency of simulation can thus be improved
up to one order of magnitude in favourable circumstances. 

The operatorial approach is very elegant and fruitful. However, it
can be cumbersome in the systematic derivation of propagators in different
contexts. A typical case is provided by non-separable Hamiltonians,
finding wide application in the description in generalized coordinates
or in presence of velocity-dependent forces (see e.g. \cite{melchionna2}).
In this case, symplectic and robust numerical schemes, such as the
Generalized Leapfrog, have been known for some time \cite{leimkuhler}.
Another wide class of dynamics is represented by stochastic equations
of motion, such as the all-famous Langevin equation. In this context,
the operatorial route has again been applied by considering the stochastic
noise as a time-dependent perturbation \cite{ricci-ciccotti} or by
evolving the state according to a Fokker-Planck propagator \cite{forbert-chin,espanol}.
It has to be mentioned that for a first-order stochastic differential
equation, a fourth-order accurate scheme has been derived \cite{forbert-chin}
based on a Runge-Kutta propagation of the deterministic component.
However, such scheme requires to compute high order derivatives, thus
being rather complicated for practical applications in condensed matter,
and is not specifically designed to reduce to symplectic in the limit
of zero friction. In general, due to the presence of both stochastic
and deterministic forces, conventional operatorial calculus should
be applied with some care\cite{kuhnemund}. As a matter of fact, a
previous operatorial approach was shown to overestimate the accuracy
of the numerical trajectory \cite{ricci-ciccotti,ciccotti-private}.

Stochastic dynamics applied to condensed matter systems has recently
received renewed attention for several reasons: i) recent emphasis
on multi-scale methods is based on the simultaneous evolution of atoms
and surrounding hydrodynamic fields, the latter been treated via Lattice
Boltzmann dynamics. In this technique noise serves to effectively
enforce fluctuation-dissipation relations in otherwise decoupled systems,
thus requiring stable and accurate numerical integrators \cite{duenweg,fyta};
ii) improving the stability of deterministic MTS algorithms to large
timesteps can be achieved by attaching to each degree of freedom massive
thermostats to overdamp the atomic motion \cite{tuckerman} or by
stabilizing the motion by an overdamped Langevin friction. The notion
of Langevin stabilization is not new, and has already been proposed
in the literature in different forms \cite{ma-izaguirre,izaguirre-catarello};
iii) Langevin dynamics is an effective mean of attaching the system
to a thermal reservoir, thus allowing to sample the canonical ensemble
without the introduction of artificial dynamics with memory (e.g.
Nosè-Hoover). 

In the present paper we will consider a deterministic class of propagators,
the Verlet being a particular case for separable Hamiltonian dynamics,
derived from a different perspective, based on splitting the phase
space vector field rather than by approximating the Liouvillean. The
trajectory representation employed here is clearly equivalent to the
operatorial one. However, in the operatorial approach one evaluates
the involved time integrals at current times (generating so-called
{}``forward'' update) while in the trajectory representation one
can evaluate such integrals with more general interpolation schemes.
The present approach allows to establish a clear connection between
the algorithm and eventual approximations supplementing the propagation
of the trajectory. In the same spirit, the technique is extended to
Langevin dynamics for both STS and MTS approaches. We will derive
numerical integrators for Langevin dynamics accurate to second-order
in the timestep and reducing to symplectic Verlet ones in the limit
of zero friction. Our derivation treats momenta as natural elements
of the algorithm, an aspect which has recently been debated \cite{wang-skeel}.
Some numerical tests demonstrate the statistical consistency of the
schemes for STS and MTS propagations. Moreover, the stochastic MTS
scheme applied to a realistic system made of water molecules, therefore
comprising intramolecular, excluded volume and electrostatic forces,
is shown to achieve improved numerical stability and correct configurational
statistics.

\section{Deterministic dynamics}

Let us consider the $6N$ phase space point $x=\{ q_{i},p_{i}\}_{i=1,3N}$
associated to the autonomous equations of motion written as

\begin{equation}
\dot{x}=S\fld(x)\label{eq:ham}\end{equation}
where $S$ is a generic matrix with constant elements and $\fld(x)$
a generic vector field. In case of Hamiltonian dynamics the vector
field reads\begin{equation}
S\fld(x)=\SS\partial H(x)\label{eq:hamfield}\end{equation}
where $\partial\equiv\{\frac{\partial}{\partial q_{i}},\frac{\partial}{\partial p_{i}}\}$.
$\SS$ is the so-called symplectic matrix 

\begin{equation}
\SS=\left(\begin{array}{cc}
0 & I_{d}\\
-I_{d} & 0\end{array}\right)\end{equation}
where the $3N\times3N$ block matrices $0$ and $I_{d}$ are the null
and identity matrix, respectively. Therefore, $\SS$ has a trivial
inverse $\SS^{-1}=\SS^{T}=-\SS$, expressing the fact that the symplectic
transformation is norm preserving ($\|\SS g\|=\| g\|$). For separable
Hamiltonians $\fld(x)$ reduces to the familiar form\begin{equation}
\partial H(x)=\left(\begin{array}{c}
-F(q)\\
p/m\end{array}\right)\label{eq:separablehamfield}\end{equation}
where $F$ and $p$ refer to the $3N$ components of forces and momenta
and $m$ are the masses, assumed to be equal for the sake of simplicity. 

%
{}

In order to integrate numerically eq.(\ref{eq:ham}), we consider
the generic decomposition \begin{equation}
S=U+L\end{equation}
where, for Hamiltonian flow, $S$ reduces to $\SS$ and $U$ and $L$
specialize to

\begin{equation}
\UU=\left(\begin{array}{cc}
0 & I_{d}\\
0 & 0\end{array}\right)~~~~~~~~\LL=\left(\begin{array}{cc}
0 & 0\\
-I_{d} & 0\end{array}\right)\end{equation}

We employ now a splitting technique to integrate the equations of
motion. The idea is to approximate the vector field (\ref{eq:ham})
as the superposition of two linearly independent vector fields, each
one associated to the equations of motion \begin{equation}
\frac{dx}{d\lambda}=L\fld(x)\end{equation}
and \begin{equation}
\frac{dx}{d\mu}=U\fld(x)\end{equation}
Symbolically, each evolution operator is briefly indicated as $\frac{d}{d\lambda}$
and $\frac{d}{d\mu}$ \cite{shutz}, where $\frac{d}{dt}$ refers
to the complete evolution (\ref{eq:ham}), associated to the updates

\begin{equation}
x_{o}\rightarrow e^{h\frac{d}{d\lambda}}x_{o}\label{eq:L}\end{equation}
and\begin{equation}
x_{o}\rightarrow e^{h\frac{d}{d\mu}}x_{o}\label{eq:U}\end{equation}
together with the composite evolutions $x_{o}\rightarrow e^{h\frac{d}{d\mu}}e^{h\frac{d}{d\lambda}}x_{o}$
and $x_{o}\rightarrow e^{h\frac{d}{d\lambda}}e^{h\frac{d}{d\mu}}x_{o}$.
The two operators $\frac{d}{d\lambda}$ and $\frac{d}{d\mu}$ do not
commute in general, i.e. the so-called Lie bracket $[\frac{d}{d\lambda},\frac{d}{d\mu}]\not=0$,
implying that the distance between the evoluted points has error of
order $h^{2}$, i.e. $e^{h\frac{d}{d\lambda}}e^{h\frac{d}{d\mu}}x_{o}=(e^{h\frac{d}{d\mu}}e^{h\frac{d}{d\lambda}}+h^{2}[\frac{d}{d\lambda},\frac{d}{d\mu}])x_{o}+O(h^{3})$
\cite{shutz}. On the other hand, by considering the symmetric composition
$e^{\frac{h}{2}\frac{d}{d\lambda}}e^{\frac{h}{2}\frac{d}{d\mu}}e^{\frac{h}{2}\frac{d}{d\mu}}e^{\frac{h}{2}\frac{d}{d\lambda}}$
this approximates the true trajectory to second-order accuracy in
a timestep $h$.

Associated with the two evolutions (\ref{eq:L},\ref{eq:U}) we now
consider either the exact solution of the dynamical equations for
each vector field, $\phi_{L,h}(x_{o})$ and $\phi_{U,h}(x_{o})$,
or the maps based on the four elementary approximations to the time
integrals $L\int_{0}^{h}\fld(t)dt$ and $U\int_{0}^{h}\fld(t)dt$
as\begin{eqnarray}
\Phi_{L,h}(x_{o}) & \equiv & x_{o}+hL\fld_{o}\nonumber \\
\bar{\Phi}_{L,h}(x_{o}) & \equiv & x_{o}+hL\fld_{h}\nonumber \\
\Phi_{U,h}(x_{o}) & \equiv & x_{o}+hU\fld_{o}\nonumber \\
\bar{\Phi}_{U,h}(x_{o}) & \equiv & x_{o}+hU\fld_{h}\label{eq:elementary-quadratures}\end{eqnarray}
where the bar denotes evaluation of the integral at the upper extremum
of the interval (requiring in principle the inversion of an implicit
equation). Conversely, the operatorial route is always based on evaluations
at the lower extremum, thus allowing for more flexibility in the design
of numerical algorithms in the present treatment. Nevertheless, the
vector field splitting and the approximated integrals are two independent
sources of numerical error. 

The Euler-A propagator is defined by the composition\begin{equation}
x_{h/2}=\Phi_{U,h/2}\circ\bar{\Phi}_{L,h/2}(x_{o})\label{eq:euler-a-quadrature}\end{equation}
and the Euler-B scheme is defined by the composition \begin{equation}
x_{h/2}=\Phi_{L,h/2}\circ\bar{\Phi}_{U,h/2}(x_{o})\label{eq:euler-b-quadrature}\end{equation}
It is easily shown that Euler-B is the adjoint of Euler-A, where the
adjoint of the map $M_{h}$ is defined as $M_{h}^{\star}=M_{-h}^{-1}$
\cite{hairer}. A symmetrized algorithm can be constructed as \begin{equation}
x_{h}=\Phi_{L,h/2}\circ\bar{\Phi}_{U,h/2}\circ\Phi_{U,h/2}\circ\bar{\Phi}_{L,h/2}(x_{o})\label{eq:EA-EB}\end{equation}
or by interchanging $U\leftrightarrow L$. Given the presence of pairs
of the type $\bar{\Phi}_{L}\circ\Phi_{L}$ and $\bar{\Phi}_{U}\circ\Phi_{U}$
in the consecutive applications of eq.(\ref{eq:EA-EB}), the global
propagator is based on evaluation of the integrals via the midpoint
method, being accurate to quadratic order in the timestep. The scheme
is symmetric with respect to time-inversion, since\begin{equation}
x_{o}=\bar{\Phi}_{L,-h/2}\circ\Phi_{U,-h/2}\circ\bar{\Phi}_{U,-h/2}\circ\Phi_{L,-h/2}(x_{h})\end{equation}

For non-separable Hamiltonians, the propagator (\ref{eq:EA-EB}) reduces
to the Generalized Leapfrog \cite{leimkuhler}. The updating scheme
reads \begin{eqnarray}
p^{\star} & = & p_{o}-\frac{h}{2}\frac{\partial H}{\partial q}(q_{o},p^{\star})\nonumber \\
q_{h} & = & q_{o}+\frac{h}{2}\left[\frac{\partial H}{\partial p}(q_{h},p^{\star})+\frac{\partial H}{\partial p}(q_{o},p^{\star})\right]\nonumber \\
p_{h} & = & p^{\star}-\frac{h}{2}\frac{\partial H}{\partial q}(q_{h},p^{\star})\end{eqnarray}
where $p^{\star}$ indicates intermediate values of the momenta.

For a separable Hamiltonian each update of the map $x_{h}=\phi_{\LL,h/2}\circ\phi_{\UU,h}\circ\phi_{\LL,h/2}(x_{o})$
becomes explicit, the resulting scheme given by the celebrated Velocity
Verlet (VV)\begin{eqnarray}
p^{\star} & = & p_{o}+\frac{h}{2}F(q_{o})\nonumber \\
q_{h} & = & q_{o}+h\frac{p^{\star}}{m}\nonumber \\
p_{h} & = & p^{\star}+\frac{h}{2}F(q_{h})\end{eqnarray}
The so-called Position Verlet propagator (PV) is similarly obtained
by composing the maps as $x_{h}=\phi_{\UU,h/2}\circ\phi_{\LL,h}\circ\phi_{\UU,h/2}(x_{o})$.

A central property of the Generalized Leapfrog and Verlet-based propagators
is symplecticity. Generally speaking, the symplectic property applies
to Hamiltonian dynamics, which obeys the relation \cite{leimkuhler}\begin{equation}
J^{T}\SS^{-1}J=\SS^{-1}\label{eq:sympl}\end{equation}
where $J$ is the Jacobian of the transformation $x(0)\rightarrow x(t)$.
This property is related to the fact that the dynamics is invariant
under canonical transformations. Conservation of measure follows directly,
 since eq.(\ref{eq:sympl}) implies $\| J\|^{2}=1$. 

An explicit calculation proves that the Euler-A scheme for Hamiltonian
flow is symplectic. In fact, \begin{equation}
J=\frac{\partial x_{h}}{\partial x_{o}}=I+h\SS\partial^{2}H_{h/2}\frac{\partial x_{h/2}}{\partial x_{o}}=I+h\SS\Omega_{h/2}(I-h\LL\Omega_{h/2})^{-1}=(I+h\LL\Omega_{h/2})(I-h\UU\Omega_{h/2})^{-1}\end{equation}
where $\Omega=\Omega^{T}\equiv\partial^{2}H$ and having using the
fact that $x_{o}=x_{h/2}-h\LL\partial H_{h/2}$, so that $\frac{\partial x_{h/2}}{\partial x_{o}}=\left(\frac{\partial x_{o}}{\partial x_{h/2}}\right)^{-1}=(I-h\LL\Omega_{h/2})^{-1}$.
Finally, \begin{eqnarray}
J^{T}\SS^{-1}J-\SS^{-1} & = & \left[I+\frac{h}{2}\SS\Omega_{h/2}(I-\frac{h}{2}\LL\Omega_{h/2})^{-1}\right]^{T}\SS^{-1}\left[I+\frac{h}{2}\SS\Omega_{h/2}(I-\frac{h}{2}\LL\Omega_{h/2})^{-1}\right]-\SS^{-1}\nonumber \\
 & = & [(I-\frac{h}{2}\LL\Omega_{h/2})^{-1}]^{T}\left[\frac{h}{2}\Omega_{h/2}\LL\Omega_{h/2}+\frac{h}{2}\Omega_{h/2}\UU\Omega_{h/2}+\frac{h}{2}\Omega_{h/2}\SS^{T}\Omega_{h/2}\right]\nonumber \\
 &  & \times(I-\frac{h}{2}\LL\Omega_{h/2})^{-1}=0\end{eqnarray}

A central property of symplectic propagators regards the presence
of a shadow Hamiltonian exactly conserved by the numerical flow and
written as a power expansion in the timestep, $\tilde{H}(x)=H(x)+\sum_{n}h^{n}G^{n}(x)$,
such that the associated vector field is equal to $\Delta x\equiv x_{h}-x_{0}=hS\partial\tilde{H}$.
An explicit expression for the leading $G^{n}(x)$ terms can be derived
with leading terms $hG^{1}(x)$ and $-hG^{1}(x)$ for symplectic Euler-A
and Euler-B schemes respectively. As a consequence, VV scheme has
leading term of $h^{2}$ order \cite{leimkuhler}. Clearly, on the
$\tilde{H}=const$ manifold, measure is preserved, since $\partial\cdot\Delta x=0$,
and one can define a shadow distribution function $\tilde{f}(y,t)$
such that a Liouville equation is exactly satisfied\begin{equation}
\tilde{f}(y,h)-\tilde{f}(y,0)+\frac{1}{h}\Delta x\partial\tilde{f}(y,0)=0\end{equation}

\section{Langevin dynamics}

Let us consider the Langevin dynamics\begin{equation}
\left(\begin{array}{c}
\dot{q}\\
\dot{p}\end{array}\right)=\left(\begin{array}{c}
p/m\\
F(q)-\gamma p+\eta(t)\end{array}\right)\label{eq:langevin-qp}\end{equation}
where $\eta(t)$ is a white noise, with\begin{eqnarray}
\langle\eta(t)\rangle & = & 0\nonumber \\
\langle\eta(t)\eta(t')\rangle & = & g^{2}\delta(t-t')\end{eqnarray}
$\gamma$ is the friction coefficient, $g^{2}\equiv2\gamma mkT$,
and the deterministic part arises from a separable Hamiltonian. The
present treatment, applied to the case of a scalar constant friction,
can be rapidly extended to the case of a position-dependent or tensorial
friction, as used in the presence of hydrodynamics or confined diffusion.
The equations of motion are written in compact notation as\begin{equation}
\dot{x}=(\SS+mG)\partial H(x)+\tilde{\eta}(t)\label{eq:langevin-compact}\end{equation}
where\begin{equation}
G=-\gamma\left(\begin{array}{cc}
0 & 0\\
0 & I_{d}\end{array}\right)\,\,\,\,\,\,\,\,\,\,\,\,\tilde{\eta}(t)=\left(\begin{array}{c}
0\\
\eta(t)\end{array}\right)\end{equation}

Different algorithms can be employed to integrate the Langevin dynamics,
the most popular being the Euler and the Heun ones \cite{kloeden-platen}
whereas, in presence of interatomic forces, different algorithms have
been proposed in the past \cite{vangunsteren-berendsen,brunger,allentildesley,skeel-izaguirre,mannella}.
The numerical solution of (\ref{eq:langevin-qp}) presents, in general,
non trivial mixing between deterministic and stochastic terms. This
is shown by a standard argument \cite{honeycutt}. Let us consider
an elementary step written in Euler form, $\delta x_{t}=f_{o}t+\tilde{W}_{t}$,
where $f(x)=(S+mG)\partial H(x)$ and $\tilde{W}_{t}=\int_{0}^{t}\tilde{\eta}(s)ds$
is the Wiener increment \cite{gardiner}. The Euler step is exact
at sufficiently small time $t$ and can now be nested in the integral
of (\ref{eq:langevin-compact}), leading to the formal solution\begin{eqnarray}
x_{h}-x_{0} & = & \int_{0}^{h}dt(f_{0}+\partial f_{0}\delta x_{t}+\frac{1}{2}\partial^{2}f_{0}\delta x_{t}^{2}+....+\tilde{\eta}_{t})=\nonumber \\
 & = & f_{0}h+\tilde{W}_{h}+\partial f_{0}f_{0}\frac{h^{2}}{2}+\partial f_{0}\int_{0}^{h}dt\tilde{W}_{t}+O(h^{5/2})\label{eq:euler-exp}\end{eqnarray}
This expression shows that, in order to obtain the desired accuracy,
a second order integrator for the deterministic component is needed
together with a proper handling of the stochastic forces. In fact,
to second order, deterministic and stochastic terms mix up via a linear
functional of the noise, $\partial f_{0}\int_{0}^{h}dt\tilde{W}_{t}$,
scaling as $h^{3/2}$. Due to the separable Hamiltonian, direct inspection
reveals that this term reduces simply to $1/m\int_{0}^{h}dtW_{t}$
for positions and $-\gamma\int_{0}^{h}dtW_{t}$ for momenta, i.e.
arising from inertial and Langevin forces only. In principle, the
purely deterministic terms ($f_{0}h+\partial f_{0}f_{0}\frac{h^{2}}{2}+...$),
accounting for the effect of the operator $e^{hf\partial}(x_{0})$,
cannot be separated from the stochastic ones via the simplistic operatorial
splitting, at the expense of missing the mixed term in eq.(\ref{eq:euler-exp}).
However, it should be possible to decompose the vector fields by separating
the evolution of the configurational degrees of freedom from that
of the momenta, by considering the splitting\begin{equation}
\dot{x}=(\LL+mG)\partial H(x)+\tilde{\eta}(t)\label{eq:L+D}\end{equation}
and\begin{equation}
\dot{x}=\UU\partial H(x)\label{eq:Udue}\end{equation}
Eq. (\ref{eq:L+D}) has solution given by the exact map\begin{eqnarray}
\psi_{\LL,h}(x_{o}) & \equiv & \left(\begin{array}{c}
q_{0}\\
e^{-\gamma h}p_{0}+\frac{(1-e^{-\gamma h})}{\gamma}F_{0}+Q_{h}\end{array}\right)\end{eqnarray}
where the following stochastic integral appears\begin{equation}
Q_{h}=e^{-\gamma h}\int_{o}^{h}dse^{\gamma s}\eta(s)=\NN(mkT(1-e^{-2\gamma h}))\label{eq:wiener}\end{equation}
with $\NN(\sigma^{2})$ being a gaussian variable with zero mean and
variance $\sigma^{2}$.

We define the stochastic Euler-A as\begin{eqnarray}
x_{h/2} & = & \phi_{\UU,h/2}\circ\psi_{\LL,h/2}(x_{o})\end{eqnarray}
and the stochastic Euler-B as\begin{eqnarray}
x_{h/2} & = & \psi_{\LL,h/2}\circ\phi_{\UU,h/2}(x_{o})\end{eqnarray}

By symmetric composition, the Stochastic Velocity Verlet (SVV) is
constructed as \begin{eqnarray}
x_{h} & = & \psi_{\LL,h/2}\circ\phi_{\UU,h}\circ\psi_{\LL,h/2}(x_{o})\end{eqnarray}
where we have collapsed $\phi_{\UU,h/2}\circ\phi_{\UU,h/2}=\phi_{\UU,h}$.
The complete updating scheme reads\begin{eqnarray}
p^{\star} & = & e^{-\gamma h/2}p_{o}+\frac{(1-e^{-\gamma h/2})}{\gamma}F(q_{o})+\NN^{(1)}(mkT(1-e^{-\gamma h}))\nonumber \\
q_{h} & = & q_{o}+\frac{h}{m}p^{\star}\nonumber \\
p_{h} & = & e^{-\gamma h/2}p^{\star}+\frac{(1-e^{-\gamma h/2})}{\gamma}F(q_{h})+\NN^{(2)}(mkT(1-e^{-\gamma h}))\end{eqnarray}
where $\NN^{(1)}$ and $\NN^{(2)}$ are two independent realizations
of the process (\ref{eq:wiener}) and thus, the algorithm requires
two extractions of random numbers per timestep. %
{}

By reverse composition, the stochastic position Verlet (SPV) is constructed,
reading \begin{eqnarray}
x_{h} & = & \phi_{\UU,h/2}\circ\psi_{\LL,h}\circ\phi_{\UU,h/2}(x_{o})\end{eqnarray}
where it is easily shown that $\psi_{\LL,h/2}\circ\psi_{\LL,h/2}=\psi_{\LL,h}$,
having collapsed the sum of independent gaussian terms into a single
gaussian extraction. In explicit terms, SPV reads\begin{eqnarray}
q^{\star} & = & q_{0}+\frac{h}{2m}p_{0}\nonumber \\
p_{h} & = & e^{-\gamma h}p_{o}+\frac{(1-e^{-\gamma h})}{\gamma}F(q^{\star})+\NN(mkT(1-e^{-2\gamma h}))\nonumber \\
q_{h} & = & q^{\star}+\frac{h}{2m}p_{h}\end{eqnarray}

%
{}

The accuracy of the SVV and SPV algorithms is evaluated below. At
first, let us consider the explicit solution of eq.(\ref{eq:langevin-qp})\begin{eqnarray}
x_{t} & = & \left(\begin{array}{c}
q_{0}+\int_{0}^{t}dt'e^{-\gamma t'}\frac{1}{m}\left[p_{0}+\int_{0}^{t'}dt''e^{\gamma t''}F(q(t''))\right]+\frac{1}{m}\int_{0}^{t}dt'e^{-\gamma t'}\int_{0}^{t'}dt''e^{\gamma t''}\eta(t'')\\
e^{-\gamma t}p_{0}+e^{-\gamma t}\int_{0}^{t}dt'e^{\gamma t'}F(q(t'))+e^{-\gamma t}\int_{0}^{t}dt'e^{\gamma t''}\eta(t')\end{array}\right)\nonumber \\
 & = & \left(\begin{array}{c}
q_{0}+\left(\frac{1-e^{-\gamma t}}{\gamma m}\right)p_{0}+\frac{1}{\gamma m}\int_{0}^{t}dt'F(q(t'))\left(1-e^{\gamma(t-t')}\right)+\frac{1}{\gamma m}(W_{t}-Q_{t})\\
e^{-\gamma t}p_{0}+e^{-\gamma t}\int_{0}^{t}dt'e^{\gamma t'}F(q(t'))+Q_{t}\end{array}\right)\label{eq:langevin-solution}\end{eqnarray}
where we used the rule $\int_{0}^{t}dt'\int_{0}^{t'}dt''=\int_{0}^{t}dt''\int_{t''}^{t}dt'$.
The variable $Q_{t}$ is easily computed to give rise to the following
covariances $\langle Q_{t}W_{t}\rangle=g^{2}(1-e^{-\gamma t})/\gamma$
and $\langle Q_{t}^{2}\rangle=g^{2}(1-e^{-2\gamma t})/2\gamma$ \cite{gardiner}.
As a result, for a constant force the covariance matrix of positions
and momenta reads\begin{eqnarray}
\langle\Delta q_{t}^{2}\rangle & = & \frac{g^{2}}{m^{2}\gamma^{3}}\left(\gamma t-2(1-e^{-\gamma t})+\frac{(1-e^{-2\gamma t})}{2}\right)=\frac{g^{2}t^{3}}{3m^{2}}+O(t^{4})\nonumber \\
\langle\Delta q_{t}\Delta p_{t}\rangle & = & \frac{g^{2}}{2m\gamma^{2}}\left(1-e^{-\gamma t}\right)^{2}=\frac{g^{2}t^{2}}{2m}(1-\gamma t)+O(t^{4})\nonumber \\
\langle\Delta p_{t}^{2}\rangle & = & \frac{g^{2}}{2\gamma}\left(1-e^{-2\gamma t}\right)\label{eq:covariances}\end{eqnarray}
We now confront these terms with the corresponding ones appearing
in SVV and SPV. SVV generates the following noise terms\begin{eqnarray}
\Delta q_{h} & = & \frac{h}{m}Q_{h/2}^{(1)}\nonumber \\
\Delta p_{h} & = & e^{-\gamma h/2}Q_{h/2}^{(1)}+Q_{h/2}^{(2)}\label{eq:SVVnoise}\end{eqnarray}
with covariances given by $\langle\Delta q_{h}^{2}\rangle=g^{2}h^{2}(1-e^{-\gamma h})/2\gamma m^{2}=g^{2}h^{3}/2m^{2}+O(h^{4})$,
$\langle\Delta q_{h}\Delta p_{h}\rangle=g^{2}h(e^{-\gamma h/2}-e^{-3\gamma h/2})/2\gamma m=g^{2}h^{2}(1-\gamma h)/2m+O(h^{4})$
and $\langle\Delta p_{h}^{2}\rangle=g^{2}(1-e^{-2\gamma h})/2\gamma$.
Similarly, for SPV we find that $\langle\Delta q_{h}^{2}\rangle=g^{2}h^{3}/4m+O(h^{4})$,
$\langle\Delta q_{h}\Delta p_{h}\rangle=g^{2}h^{2}(1-\gamma h)/2m+O(h^{4})$
and $\langle\Delta p_{h}^{2}\rangle=g^{2}(1-e^{-2\gamma h})/2\gamma$.
In conclusion, for both SVV and SPV positional variance coincides
with (\ref{eq:covariances}) up to quadratic order, position-momentum
covariance up to cubic order, and momentum to any order, of the form
sought. 

Before proceeding further, we wish to make a couple of comments regarding
the derived schemes. The first is that in numerical applications,
one is usually concerned with the usage of uniform random number generators
in lieu of the expensive Guassian ones. This is typically possible
for the simple Euler scheme \cite{duenweg-paul} in which a uniform
variate is used to sample a gaussian process up to second moment.
In principle, a second order accuracy in the propagator would require
the usage of a linear combination of uniform variates in order to
sample the gaussian distribution up to the forth moment. In our treatment,
we have used the information up to the second moment of the stochastic
processes $Q_{t}$ or $W_{t}$. This circumstance, mirrored by the
presence of two random variables arise in the Verlet-like class of
integrators, allows to employ uniform random number generators in
the present case. The second comment regards the SVV scheme, whose
integration of the deterministic part has already appeared in the
literature \cite{allen82}, and shown to be equivalent to the one
proposed by van Gunsteren and Berendsen \cite{vangunsteren-berendsen}.
In the present approach, however, the noise terms are not imposed
to be equal to the matrix (\ref{eq:covariances}), as in previous
approaches, but rather emerge spontaneously, ultimately due to the
decomposition of the underlying vector field.

The derivation of SVV and SPV is based on the exact integration of
eqs. (\ref{eq:L+D},\ref{eq:Udue}). However, one could integrate
the decomposed dynamics by evaluating the time integrals with expressions
analogous to (\ref{eq:elementary-quadratures}), but for the stochastic
case. In particular, by defining the maps \begin{eqnarray}
\Psi_{\LL,h}(x_{o}) & \equiv & \left(\begin{array}{c}
q_{0}\\
-\gamma hp_{0}+hF_{0}+W_{h}\end{array}\right)\nonumber \\
\bar{\Psi}_{\LL,h}(x_{o}) & \equiv & \left(\begin{array}{c}
q_{0}\\
-\gamma hp_{h}+hF_{0}+W_{h}\end{array}\right)\label{eq:approxStoch}\end{eqnarray}
and the compositions $\bar{\Psi}_{\LL,h/2}\circ\phi_{\UU,h}\circ\Psi_{\LL,h/2}$
or $\phi_{\UU,h/2}\circ\bar{\Psi}_{\LL,h/2}\circ\Psi_{\LL,h/2}\circ\phi_{\UU,h/2}$
another pair of Verlet-like algorithms is derived. The updating scheme
analogous to SVV, which we name SVVm, then reads \begin{eqnarray}
p^{\star} & = & \frac{1}{1+\gamma h/2}\left\{ p_{o}+\frac{h}{2}F(q_{o})+\NN^{(1)}(mkT\gamma h)\right\} \nonumber \\
q_{h} & = & q_{o}+\frac{h}{m}p^{\star}\nonumber \\
p_{h} & = & (1-\frac{\gamma h}{2})p^{\star}+\frac{h}{2}F(q_{h})+\NN^{(2)}(mkT\gamma h)\end{eqnarray}
A straightforward calculation shows that the corresponding moments
are again accurate to second order with the ones of eq.(\ref{eq:covariances}). 

It is easily verified that the deterministic VV and PV propagators
are recovered from SVV and SPV (or the approximate SVVm) in the limit
$\gamma\rightarrow0$. If morever the Jacobian is phase-space independent,
the method is called quasi-symplectic \cite{milstein-tretyakov}.
Let us consider, as an example, the stochastic Euler-A method,\begin{eqnarray}
\frac{\partial x_{o}}{\partial x_{h/2}} & = & e^{-Gh/2}+\left(\frac{e^{\gamma h/2}-1}{\gamma}\right)\LL\Omega_{h/2}\end{eqnarray}
and\begin{equation}
J=\frac{\partial x_{h}}{\partial x_{o}}=\left(\frac{\partial x_{h}}{\partial x_{h/2}}\right)\left(\frac{\partial x_{o}}{\partial x_{h/2}}\right)^{-1}=\left(I+\frac{h}{2}\UU\Omega_{h/2}\right)\left(e^{-Gh/2}-\left(\frac{e^{\gamma h/2}-1}{\gamma}\right)\LL\Omega_{h/2}\right)^{-1}\end{equation}
so that, for a block diagonal $\Omega$, $\| J\|=e^{-3N\gamma h/2}$.
The same result is derived for the stochastic Euler-B scheme. Globally,
SVV and SPV have a Jacobian given by$\| J\|=e^{-3N\gamma h}$. Moreover,
this property allows to employ the propagators in a Hybrid Monte Carlo
scheme, based on an underlying Langevin dynamics. 

The statistical distribution associated to the numerical Langevin
flow map is readily derived. Its evolution is given by \begin{equation}
\tilde{f}(y,t+h)-\tilde{f}(y,t)=\int dx\left[\delta(y-x_{t}-\Delta x)-\delta(y-x_{t})\right]\tilde{f}(x_{t},t)=\int dx\sum_{n=1}^{\infty}\frac{(-1)^{n}}{n!}\delta(y-x_{t})\partial^{n}\left(\langle\Delta x^{n}\rangle_{\eta}\tilde{f}(x,t)\right)\label{eq:LangevinEvolution}\end{equation}
having Taylor expanded the delta function and integrated by parts.
Moreover, $\langle\cdot\rangle_{\eta}$ indicates averaging over noise.
By considering a first order scheme, e.g. the stochastic Euler A,
and up to a first order dependence in the timestep, we write $\Delta x=\Delta x_{H}+\Delta x_{\gamma}$,
where $\Delta x_{H}$ is the variation due to Hamiltonian dynamics
(at $\gamma=0$), being the vector field arising from the shadow Hamiltonian
$\tilde{H}$ of the corresponding Hamiltonian numerical scheme. Moreover,
$\Delta x_{\gamma}$ arises from the dissipative and random terms.

By retaining the first two terms of the right hand side of eq. (\ref{eq:LangevinEvolution})
we arrive at the following evolution equation\begin{equation}
\tilde{f}(y,t+h)-\tilde{f}(y,t)=-\frac{1}{h}\langle\Delta x_{H}\rangle_{\eta}\partial\tilde{f}(y,t)-\partial\left\{ G\left[\frac{1}{h}\langle\Delta x_{\gamma}\rangle_{\eta}+mk\tilde{T}\partial\right]\right\} \tilde{f}(y,t)\label{eq:LiouvilleFokkerPlanck}\end{equation}
where we used the fact that, according to standard Langevin analysis,
$\langle\Delta p^{2}\rangle_{\eta}=mk\tilde{T}h$ which dominates
over $\langle\Delta q^{2}\rangle_{\eta}\sim h^{2}$ and $\tilde{T}$
is an effective numerical temperature associated to the momentum fluctuations.
Moreover, in deriving (\ref{eq:LiouvilleFokkerPlanck}) explicit use
of the symplectic character of the Hamiltonian flow has been used.
Equation (\ref{eq:LiouvilleFokkerPlanck}) represents the numerical
Liouville equation coupled to a Fokker-Planck evolution, with stationary
solution $\tilde{f}\propto e^{-\tilde{H}/k\tilde{T}}$. 

For a second order quasi-symplectic numerical scheme, the information
gained from the elementary building blocks (Euler A and B) can be
used and at equilibrium the Boltzmann distribution is recovered up
to order $h^{2}$. In this case, the Boltzmann factor contains the
shadow Hamiltonian of the underlying scheme at $\gamma=0$.

We perform some numerical tests primarily aimed at controlling the
quality of the integrators and the convergence of the first two moments
to the theoretical expectations. An elementary test is provided by
the one-dimensional stochastic oscillator, with potential $U(q)=q^{2}/2$,
integrated in a single timestep scheme with the SVV and SPV algorithms.
We use $h=0.1$, $\gamma=0.1$ and $T=0.5$. The timestep is well
below the stability limit $h=2$, valid for the purely deterministic
VV method. The computed momentum and configurational distributions,
$P_{sim}(p,t)$ and $P_{sim}(q,t)$, should converge towards the theoretical
forms, $P(p,\infty)\sim e^{-\beta p^{2}/2}$ and $P(q,\infty)\sim e^{-\beta U(q)}$,
respectively. The rate of convergence of the normalized histograms
is monitored by the norms\begin{eqnarray}
E_{p}(T) & = & \alpha_{p}\max_{i\in N}|P_{sim}(p_{i},T)-P(p_{i},\infty)|\nonumber \\
E_{q}(T) & = & \alpha_{q}\max_{i\in N}|P_{sim}(q_{i},T)-P(q_{i},\infty)|\label{eq:norms}\end{eqnarray}
where $N$ is the number of bins and $\alpha_{q}$ and $\alpha_{p}$
are arbitrary factors, as a function of the sampling time window $T$.
In Fig. \ref{cap:Histo} the histograms for the SVV motion are shown,
and similar profiles are found for the SPV case. The plot shows that
the kinetic and configurational moments are correctly sampled down
to the distant tails of the distributions. The rate of convergence
(Fig. \ref{cap:Convergence}) is systematic, and faster for the configurational
counterpart. 

We next follow the momentum and position second moments (i.e. the
so-called kinetic and configurational temperatures) in the high timestep/high
friction limit. In this circumstance, it is often reported that Langevin
integrators produce distinct and systematic departures of kinetic
and configurational temperatures from the present input, such that
equipartion is violated, i.e. $\langle q^{2}\rangle\not=\langle p^{2}\rangle\not=kT$
\cite{allen}. Here, we employ the information from the underlying
Hamiltonian propagator to analyze such behavior and showing that,
given the quasi-symplectic form of the integrator, very accurate results
can be obtained as compared to non quasi-symplectic ones. The VV update
reads\begin{eqnarray}
q_{h} & = & (1-\frac{h^{2}}{2})q_{0}+hp_{0}\nonumber \\
p_{h} & = & (1-\frac{h^{2}}{2})p_{0}-h(1-\frac{h^{2}}{4})q_{0}\end{eqnarray}
showing that, to second order, the dynamics arises from a shadow Hamiltonian
of the form $\tilde{H}=\frac{1}{2}p^{2}+\frac{1}{2}(1-\frac{h^{2}}{4})q^{2}$.
In spite of the symmetry of the equations of motion in the $q,p$
variables, the fact that the discretized evolution presents a biased,
reduced force constant is apparently odd. Moreover, if now the dynamics
is equipped with the Langevin thermostat, it is understood that equipartition
is violated, i.e. positional fluctuations deviate quadratically from
the kinetic ones \cite{allen}. 

However, if we now consider the PV algorithm, constructed via a shift
of the updating algorithm by half timestep with respect to the VV
one, the update reads\begin{eqnarray}
q_{h} & = & (1-\frac{h^{2}}{2})q_{0}+h(1-\frac{h^{2}}{4})p_{0}\nonumber \\
p_{h} & = & (1-\frac{h^{2}}{2})p_{0}-hq_{0}\end{eqnarray}
showing that, by a simple shift of the {}``observation'' time, the
mass of the particle is rescaled by the same factor $(1-h^{2}/4)$.
Such factor is a manifestation of the symmetry of the original continuous
dynamics, although it shows up at shifted times as compared to the
VV case. If the Langevin thermostat is added, one observes a systematic
shift of the kinetic temperature from the input one at increasing
timesteps. The outcome of the present argument is that in order to
obtain equilibrated observables, functions of position or velocity
separately, one should samples these quantities at unequal times,
representing optimal sampling points along the trajectory. 

In Fig. \ref{fig:harm} we plot the configurational and kinetic temperature
of the harmonic oscillator sampled at mid ($(m+\frac{1}{2})h$, $m=0,1,2...$)
and full ($mh$, $m=0,1,2...$) time steps. The data show that SVV
produces excellent averages for both kinetic and configurational temperature
at high timesteps, only if these are sampled at unequal times. It
is important to notice that such averages are extremely robust, in
fact they keep close to the theoretical value up to timestep $h=2$.
Raising friction has the effect of shifting configurational temperature
at mid-steps by about 10\%, while kinetic temperature at full-steps
weakly deviates from the input value. In conclusion, at moderate friction
one has a way to sample well equilibrated quantities since the ballistic
behavior is well under control, while for increasing $\gamma$ systematic
errors appears in the fluctuations.

%
{}

The accuracy of the SVV, SPV and SVVm integrators to sample the dynamical
evolution is checked for the same stochastic harmonic oscillator for
$kT=1$ and $\gamma=0.1$. Given the initial condition $q(0)=p(0)=0$,
the second moments are given by \cite{chandresekhar}\begin{eqnarray}
\langle q^{2}(t)\rangle-\langle q(t)\rangle^{2} & = & 1+\frac{e^{-t}}{3}[-4+\cos\sqrt{3}t-\sqrt{3}\sin\sqrt{3}t]\nonumber \\
\langle p^{2}(t)\rangle-\langle p(t)\rangle^{2} & = & 1+\frac{e^{-t}}{3}[-4+\cos\sqrt{3}t+\sqrt{3}\sin\sqrt{3}t]\label{eq:chandresekar}\end{eqnarray}
In Fig. \ref{cap:moments} we report the numerical relative error
of the moments from the theoretical expectation (\ref{eq:chandresekar})
for times $\gg h$. The convergence of results towards the theoretical
values is apparent and follows a $h^{2}$ dependence in all cases,
qualifying the methods as weakly second order accurate.

\section{Multiple time step dynamics}

In this section, the multiple time step splitting extensions for both
the deterministic and stochastic dynamics are derived, both with underlying
separable Hamiltonian. We consider the decomposition of the Hamiltonian
into slow and fast components as\begin{equation}
\partial H(x)=\partial H^{s}(x)+\partial H^{f}(x)\end{equation}
The standard practice in MTS algorithms is to integrate the slow and
fast components separately, with timesteps $h$ and $h/n$, where
$n$ is a positive integer ($n>1$), respectively. A convenient decomposition
is given by\begin{equation}
\partial H^{s}=\left(\begin{array}{c}
-F^{s}(q)\\
0\end{array}\right)~~~~~~~~~\partial H^{f}=\left(\begin{array}{c}
-F^{f}(q)\\
p/m\end{array}\right)\end{equation}
where $F^{s}(q)$ and $F^{s}(q)$ are the slow and fast components
of the interatomic forces. The MTS version of the VV algorithm is
given by\begin{equation}
x_{h}=\phi_{\LL,h/2}^{s}\circ\left[\phi_{\LL,h/2n}^{f}\circ\phi_{\UU,h/n}\circ\phi_{\LL,h/2n}^{f}\right]^{n}\circ\phi_{\LL,h/2}^{s}(x_{o})\end{equation}
best known as the RESPA algorithm \cite{respa}, which is again measure
preserving and second order accurate in time. An equivalent MTS version
of the position Verlet propagator can be constructed along similar
lines.

The stochastic version of the MTS algorithm is obtained by associating
the frictional and noise Langevin forces to the fast part of the mechanical
forces. For example, the MTS version of the propagator generalizing
the SVV method, reads\begin{equation}
x_{h}=\phi_{\LL,h/2}^{s}\circ\left[\psi_{\LL,h/2n}^{f}\circ\phi_{\UU,h/n}\circ\psi_{\LL,h/2n}^{f}\right]^{n}\circ\phi_{\LL,h/2}^{s}(x_{o})\end{equation}

We evaluate the performance of the MTS-SVV scheme by considering a
one-dimensional model provided by a stochastic oscillator with potential
energy $U(x)=4.5x^{2}+0.025x^{4}$. The quartic term of the potential
is associated to the slow forces. The timesteps are $\pi/3$ and $\pi/30$
for the slow and fast propagators respectively. Friction is set to
$1.0$ and $0.001$ and temperature is set to $0.5$. For this choice
of the timesteps, the deterministic MTS-VV is known to produce a resonance
phenomenon \cite{chin,tuckerman}, an effect that limits its applicability
in more complex situations, such as in presence of intramolecular
bond stretching motion.

Fig. \ref{cap:MTS-quartic} illustrates the convergence towards the
expected probabilities of the configurational and momentum distributions
monitored by the norms (\ref{eq:norms}). The results exhibit a systematic
convergence for all values of friction chosen. Despite its simplicity,
the test demonstrates the consistency in the definition of momenta
even in a MTS type of propagation.

As a more realistic test, we consider a system made of 256 water molecules
and modelled by the flexible TIP3P force field (see ref.\cite{melchionna2}
for details on parameters). Following \cite{marchiprocacci}, the
forces are splitted into four different levels, although other choices
could be made \cite{qian-schlick}. The first, innermost level handles
the fast intramolecular bond stretching and angular forces. The second,
third and fourth levels integrate the short-range forces arising from
the Lennard-Jones and direct-space Coulomb interactions (via the Ewald
technique \cite{frenkel-smit}) in cut-off regions of $[0,5]$, $[5,7.5]$
and $[7.5,9.5]$ \AA{}, respectively. Moreover, the fourth level
handles the reciprocal space part of the Ewald interactions. Friction
is set to $\gamma=1~ps^{-1}$ and $\gamma=10^{3}~ps^{-1}$. Temperature
is set to $T=300~K$. The simulations are run with timesteps $(h_{o},n_{1}h_{o},n_{2}h_{o},n_{3}h_{o})$
associated to the four sub-propagators, by setting $h_{o}=0.25~fs$
for the fast propagation and choosing for the integers $(n_{1},n_{2},n_{3})$
the values $(1,1,1)$ (STS), $(8,1,1)$, $(8,2,2)$ and $(8,2,4)$.
All simulations are run for $20~ps$ total time.

In Fig. \ref{cap:gr-water} the three radial distribution functions
for the oxygen-oxygen, oxygen-hydrogen and hydrogen-hydrogen atoms
are reported for the $(1,1,1)$ and the $(8,2,2)$ runs with $\gamma=1~ps^{-1}$.
All data converge to the correct profiles, i.e. the $(1,1,1)$ run,
and equal to a set of independent simulations made by using the purely
deterministic Nosè-Hoover thermostat. However, the $(8,2,4)$ data
present distorted profiles for $\gamma=1~ps^{-1}$. A closer inspection
shows that the kinetic temperature of the system is larger than the
input value by approximately $30\%$, probably due to the large timestep
associated to the reciprocal term of electrostatics. However, by simply
increasing friction to $\gamma=10^{2}~ps^{-1}$, the simulation temperature
approaches the input value, with relative difference being less than
$5\%$, basically removing the distortion in the radial profiles.
It should be noticed that with the increased friction we maintain
the condition $\gamma h<2$ for each level of integration, which ensures
that the configurational temperature remains close to the kinetic
one. In conclusion, some overdamping of the dynamics is capable of
reducing spurious temperature shifts and configurational bais together
with stabilizing the separate propagators of the MTS scheme. A separate
study should be undertaken to investigate more systematically the
application of the Langevin MTS approach in the simulation of condensed
matter systems.

\section{Conclusions}

The present paper described the integration of Hamiltonian and Langevin
dynamical equations guided by a symplectic decomposition of the underlying
vector field in phase space. In the purely deterministic case, taken
together with appropriate quadrature formulae, the scheme provided
the basis for the Verlet and Leapfrog family of numerical propagators,
which are second-order accurate and applicable to general deterministic
(e.g. non-separable Hamiltonian) dynamics. 

In presence of stochastic forces the approach maintains a similar
splitting of the underlying Hamiltonian dynamics. The vector field
route proved convenient for the analytical treatment and we derived
numerical propagators which are weakly second order accurate. The
correctness of the configurational and momentum averages and their
fluctuations was demonstrated in a series of numerical tests. The
deterministic and stochastic propagators are rapidly extended to multiple
time step dynamics, as often employed in the simulation of condensed
matter systems. In deterministic multiple time step algorithms the
presence of resonance phenomena limits the stability range of the
methodology. By studying a realistic system composed by water, and
adopting a standard stochastic stabilization with large timesteps,
we found that a time-discretization error, manifesting itself as a
spurious shift in temperature, was reduced by employing a rather large
value of the friction coefficient. 

The performances of the proposed schemes to treat more complex stochastic
equations, such as dissipative particle dynamics \cite{hoogerbrugge-koelman},
where dissipative and random forces depend in a non trivial way on
momenta and positions, will be described in a forthcoming paper.

\newpage

\section*{Captions}

\begin{figure}[H]

\caption{\label{cap:Histo}Histograms $P(q,T)$ and $P(p,T)$ in log-linear
scale as a function of $z=q^{2}$ and $z=p^{2}$ , respectively. Results
generated with SVV for $T=10^{7}h$ simulation time. The x-axis has
been rescaled by an arbitrary factor. Squares and circles refer to
momentum and position distributions, respectively.}
\end{figure}

\begin{figure}[H]

\caption{\label{cap:Convergence}Norms $E_{p}$ and $E_{q}$ as a function
of the sampling time window for the SVV algorithm. The solid line
refers to $E_{q}$ and the dashed line refers to $E_{p}$.}
\end{figure}

\begin{figure}[H]

\caption{\label{cap:moments}Relative error of second moments from the theoretical
expectation (\ref{eq:chandresekar}) as a function of the timestep
sampled for $t=1$ and $q(0)=p(0)=0$. Sampling time is $10^{8}$
for each timestep chosen. Symbols refer to SVV data for positions
(open circles) and momenta (filled circles), SPV data for positions
(open squares) and momenta (filled squares), SVVm data for positions
(open diamonds) and momenta (filled diamonds).The dashed line corresponds
to $h^{2}$ and the dot-dashed line to $h$ trends.}
\end{figure}

\textbf{}%
\begin{figure}[H]

\caption{\label{fig:harm} Deviation of configurational and kinetic temperatures
from the input temperature for the stochastic harmonic oscillator
propagated with SVV at $kT=1$. Symbols: configurational data sampled
at mid-step (circles), and full-step (stars), and kinetic data at
mid-step (squares) and full-step (triangles). Filled symbols refer
to $\gamma=0.1$ and open symbols to $\gamma=1.0$.}
\end{figure}

\begin{figure}[H]

\caption{\label{cap:MTS-quartic}Convergence towards the expected distribution
for the SVV algorithm in the single (thick lines) and multiple (thin
lines) timestep formulations. Solid and dashed lines refer to configurational
and momentum data, respectively.}
\end{figure}

\begin{figure}[H]

\caption{\label{cap:gr-water}Radial distribution functions for water for
the (1,1,1) (solid curves) and (8,2,2) (dashed curves) simulations
with the MTS-SVV integrator (see text for details). The curves refer
to oxygen-oxygen, oxygen-hydrogen and hydrogen-hydrogen profiles.}
\end{figure}

\newpage

\begin{center}
\includegraphics[scale=0.6]{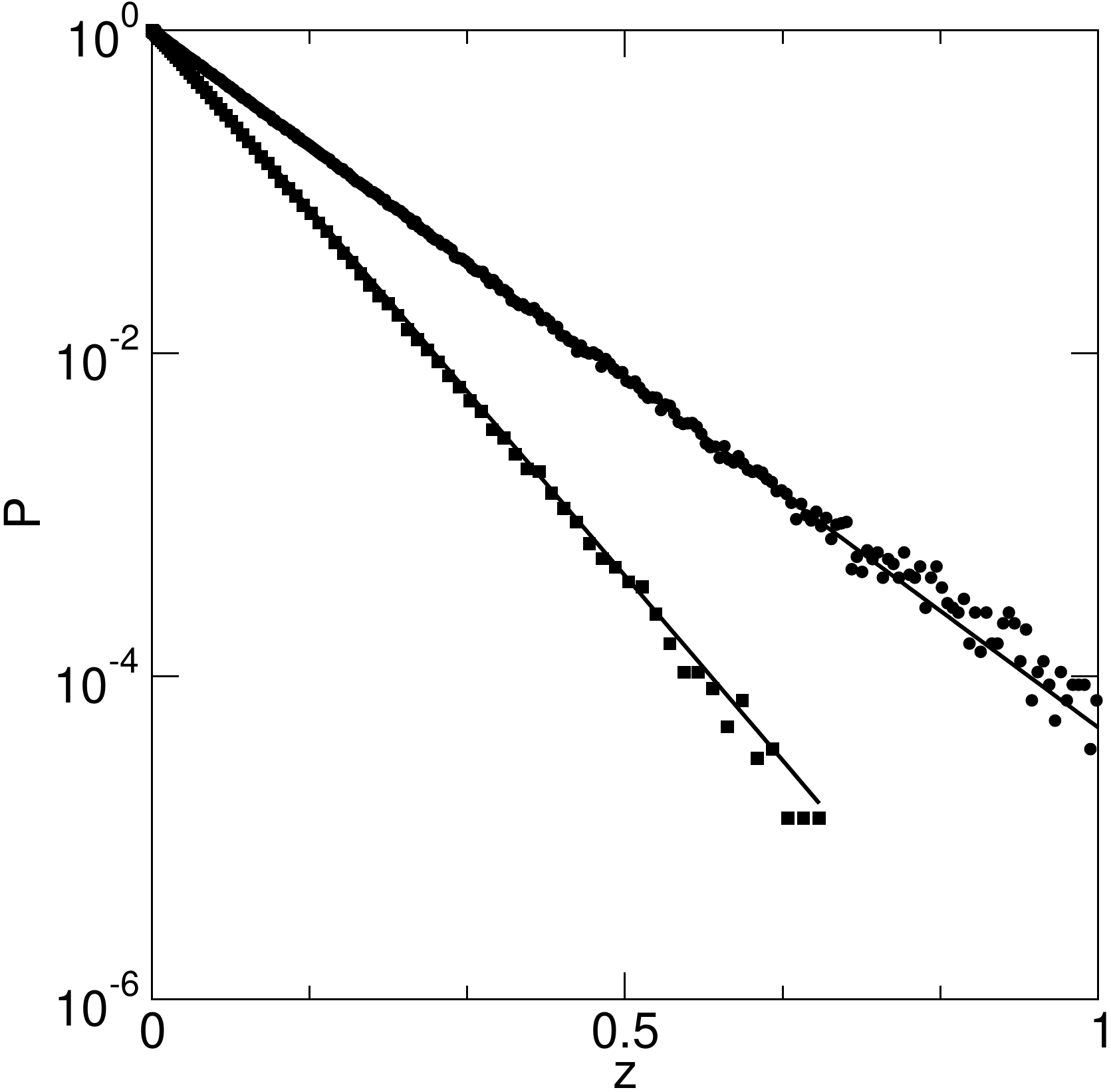}
\par\end{center}

\vspace{1cm}

Melchionna, Fig.1

\newpage

\begin{center}
\includegraphics[scale=0.6]{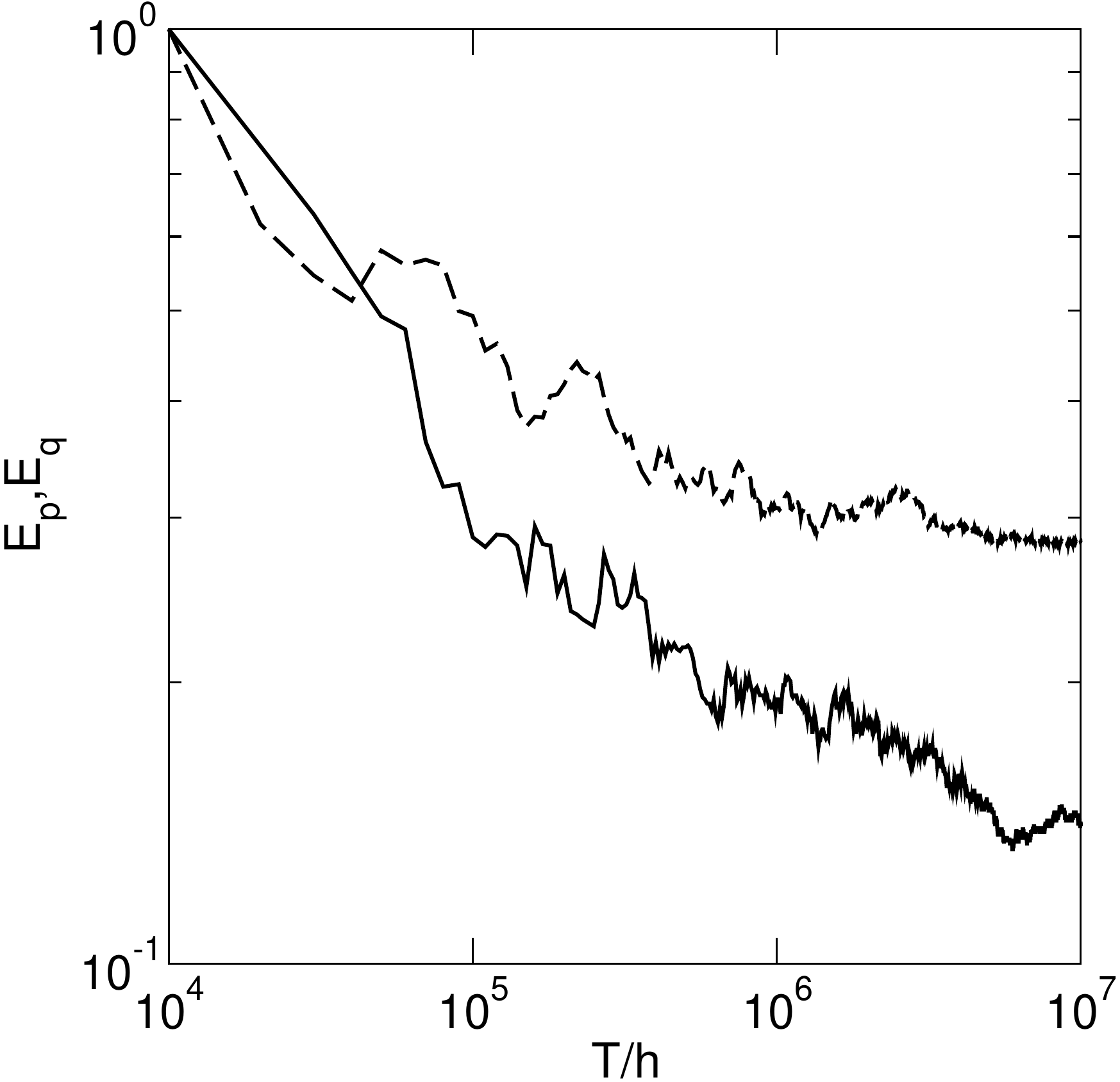}
\par\end{center}

\vspace{1cm}

Melchionna, Fig.2

\newpage

\begin{center}
\includegraphics[scale=0.6]{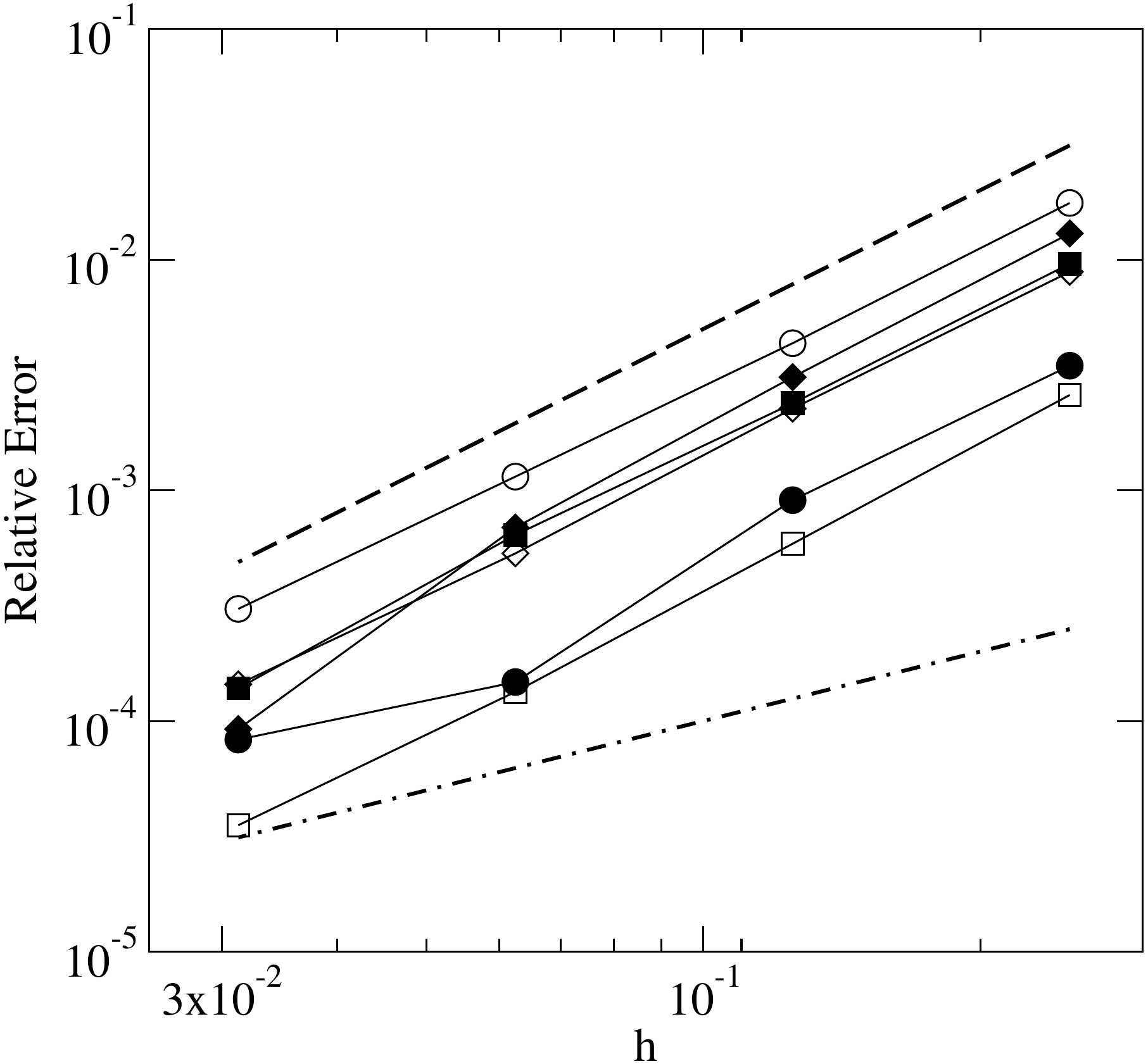}
\par\end{center}

\vspace{1cm}

Melchionna, Fig.3

\newpage

\begin{center}
\textbf{\includegraphics[scale=0.6]{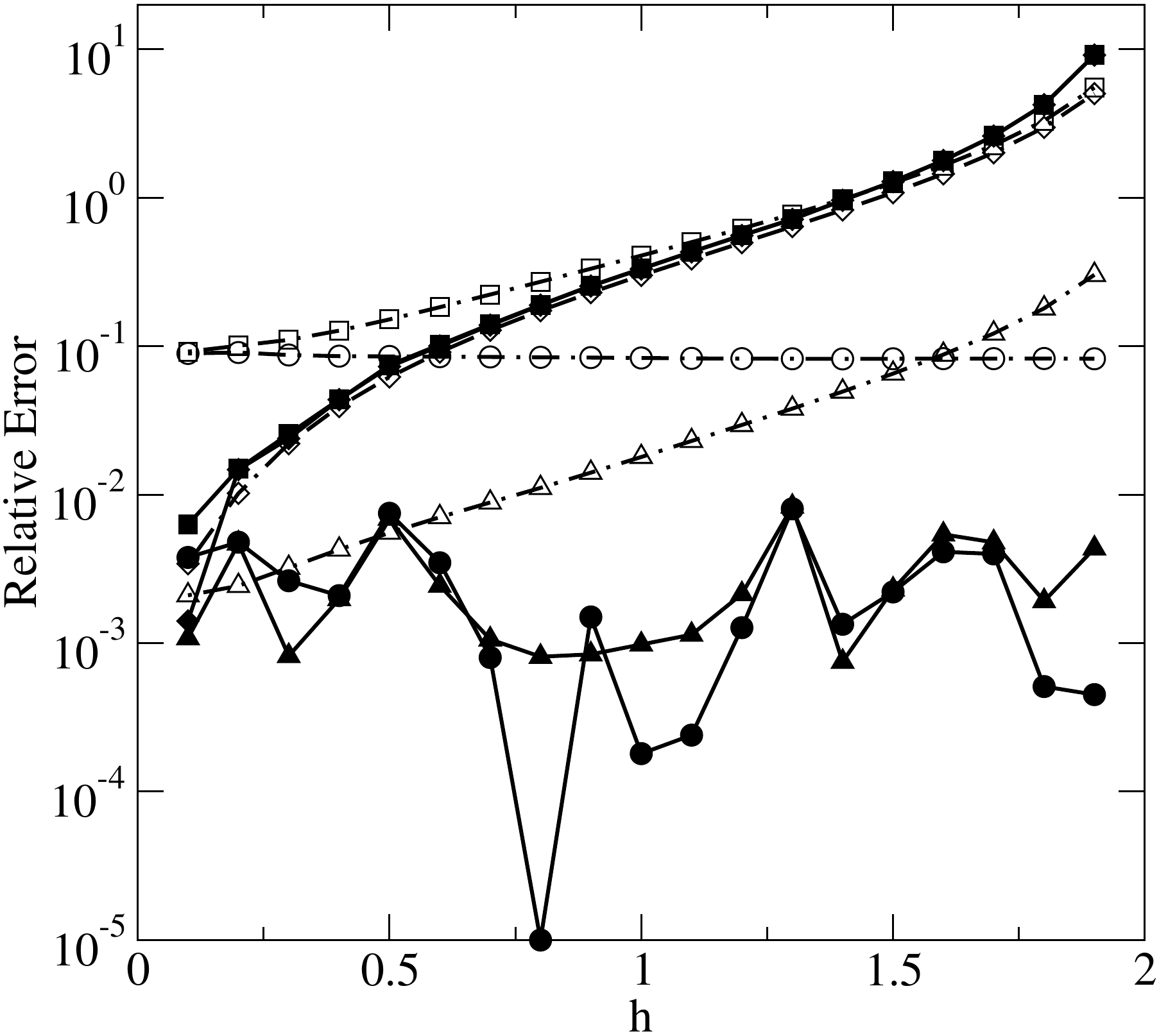}}
\par\end{center}

\vspace{1cm}

Melchionna, Fig.4

\newpage

\begin{center}
\includegraphics[scale=0.6]{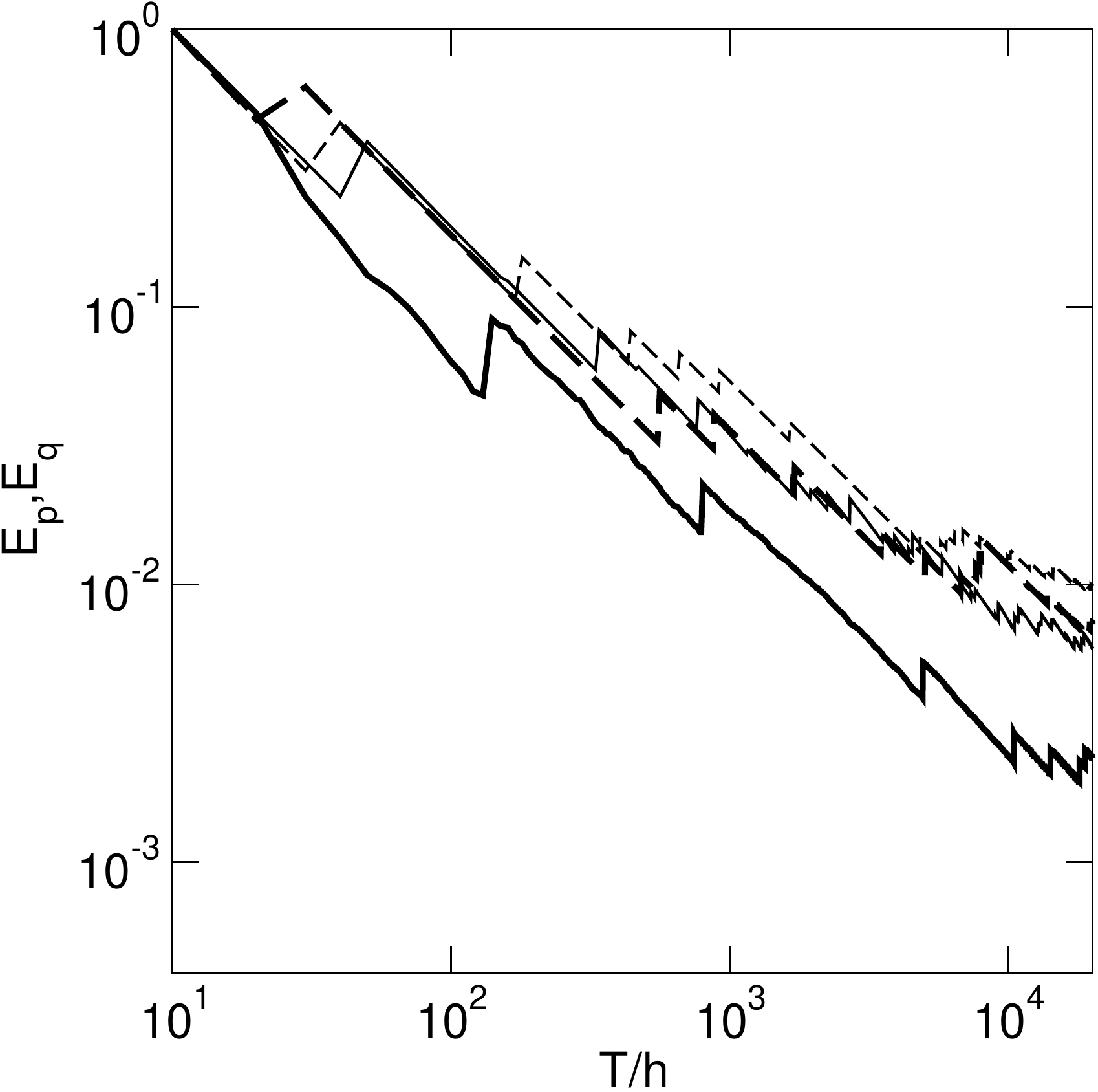}
\par\end{center}

\vspace{1cm}

Melchionna, Fig. 5

\newpage

\begin{center}
\includegraphics[scale=0.6]{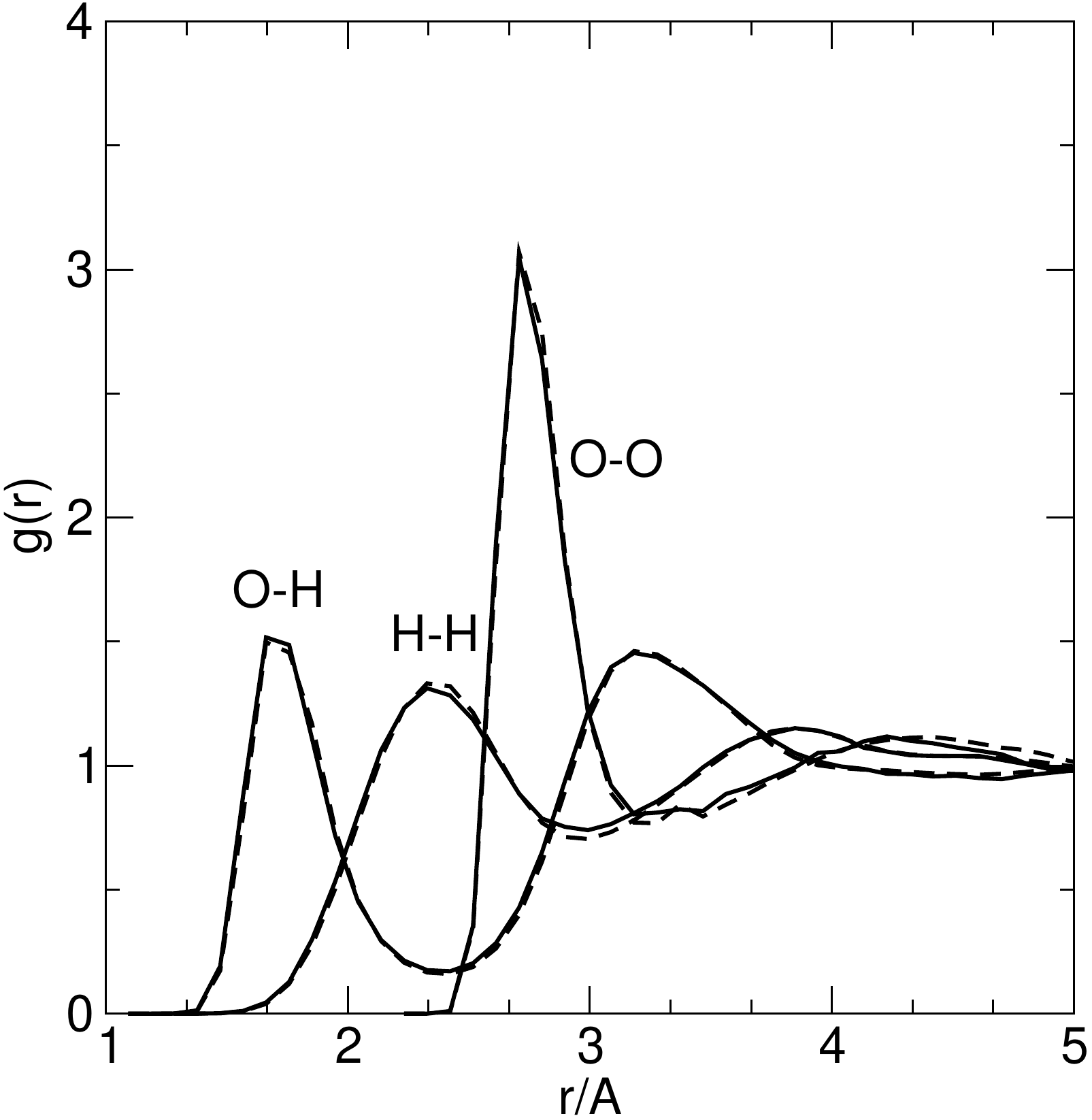} 
\par\end{center}

\vspace{1cm}

Melchionna, Fig.6
\end{document}